\documentclass[11pt]{revtex4}

\usepackage{graphicx,amsthm,amsmath, amsfonts}
\usepackage{epsfig}
\usepackage{amssymb}

\newcounter{fig}

\begin{document}


\vskip 0.cm
\rightline{\hskip 3cm ITP-UU-13/06, SPIN-13/04}
\vskip 0.cm

 \title{\Large Symmetry breaking in de Sitter: 
   \\    a stochastic effective theory approach}

\author{\Large 
   Gianrocco Lazzari\footnote{E-mail: \tt g.lazzari@students.uu.nl; t.prokopec@uu.nl}}

\author{Tomislav Prokopec$^*$}

\affiliation{Institute for Theoretical Physics, Utrecht University,\\
Postbus 80.195, 3508 TD Utrecht, The Netherlands}

\begin{abstract}

\noindent

 We consider phase transitions on (eternal) de Sitter in
an $O(N)$ symmetric scalar field theory. Making use of 
Starobinsky's stochastic inflation we prove that
deep infrared scalar modes cannot form a condensate 
-- and hence they see an effective potential that allows no phase transition.
We show that by proving convexity of the effective potential that governs deep infrared field 
fluctuations both at the origin as well as at arbitrary values of the field. 
Next, we present numerical plots of the
scalar field probability distribution function (PDF) 
and the corresponding effective potential for several values of
the coupling constant at the asymptotic future timelike infinity of de Sitter.
 For small field values the effective potential has an approximately
quadratic form, corresponding to a positive mass term, 
such that the corresponding PDF is approximately Gaussian.
However, the curvature of the effective potential shows qualitatively different 
(typically much softer) behavior on the coupling constant than that implied 
by the Starobinsky-Yokoyama procedure.
For large field values, the effective potential as expected reduces
to the tree level potential plus a positive correction
that only weakly (logarithmically) depends on the background field. 
Finally, we calculate the backreaction of fluctuations on the background geometry
and show that it is positive.

\end{abstract}


\maketitle


 \section{Introduction}
 \label{Introduction}

 The physics of de Sitter space is hard because there is no simple perturbative expansion parameter.
In particular, massless scalars and gravitons exhibit so much particle production that their 
interactions cannot be perturbatively controlled. This breakdown of perturbative expansion is a 
serious obstacle to progress of our understanding of the physics of de Sitter space. 
The usual resummation techniques do not typically help. One such resummation is 
the self-consistent Hartree (mean field) approximation~\cite{Prokopec:2011ms},
which includes the one-loop resummation of daisy (and superdaisy) diagrams. 
When applied to an O(N) symmetric real scalar field on de Sitter space, 
this approximation scheme erroneously predicts that the O(N) symmetry gets 
completely broken in the vacuum, and the (would-be) Goldstone bosons acquire a mass~\cite{Prokopec:2011ms}. 
To get this result it is sufficient to assume the self-consistent Hartree and the de Sitter symmetry. 
A notable observation is that the mass of the Goldstones is strictly smaller than the mass of 
the condensate field. Recently, a more sophisticated resummation scheme based on 
a large $N$ expansion in an $O(N)$ symmetric model has been utilized
in Refs.~\cite{Serreau:2013koa,Serreau:2013psa} 
to study deep infrared correlators on de Sitter space.
Furthermore, based on the Euclidean approach to de Sitter, 
progress has been also made towards understanding mass generation
in an interacting scalar field theory 
with a quartic self-interaction and one real scalar field 
(the $O(1)$ model)~\cite{Rajaraman:2010xd,Beneke:2012kn}.
Earlier resummation attempts 
include~\cite{Riotto:2008mv,Burgess:2009bs,Burgess:2010dd,Serreau:2011fu,Garbrecht:2011gu,Boyanovsky:2012nd}.

 In this work we make use of a sophisticated resummation technique on de Sitter space
known as stochastic inflation~\cite{Starobinsky:1986fx,Starobinsky:1994bd}. 
Stochastic inflation provides a clever reorganization and resummation of perturbation theory 
in such a way that one obtains the correct leading order answers
for infrared (super-Hubble) field correlators. 
The main observation due to Starobinsky~\cite{Starobinsky:1986fx} is that, 
while an interacting quantum field theory is essentially quantum on sub-Hubble
length scales, it is classical on super-Hubble length scales if it couples non-conformally
to gravity.
Indeed, upon splitting the theory into short and long wavelenth modes, one finds that the dynamics of long wavelength modes is particularly simple: the modes with a super-Hubble wavelength exhibit overdamped dynamics,
due to Universe's expansion gradient terms can be dropped and the only coupling between 
different modes comes from interactions. The coupling between the short and long 
wavelength modes can be modeled as a Markowian random force in the equation for the 
infrared modes. The resulting classical stochastic theory is particularly simple, 
and it can be shown to be equivalent to a Fokker-Planck equation
for the single field probability distribution function (PDF) $\rho=\rho(\phi(\vec x),t)$,
which is of the form, 
\begin{equation}
 \partial_t \rho = \frac{1}{3H}\partial_\phi\left(V^\prime \rho\right) + \frac{H^3}{8\pi^2}\partial_\phi^2\rho
\,.
 \label{Fokker-Planck}
 \end{equation}
This equation is the stochastic equivalent of the von Neumann equation for the density 
operator in quantum field theory, and thus $\rho(\phi,t)$ can be thought of
as the classical limit of the density operator. 

 Furthermore, it is known that a suitably adapted stochastic 
theory of inflation captures correctly 
field correlators in theories such as a self-interacting real 
scalar field theory~\cite{Tsamis:2005hd}, Yukawa theory~\cite{Miao:2006pn}
and scalar electrodynamics~\cite{Prokopec:2007ak,Prokopec:2008gw,Prokopec:2006ue}.
Due to the complex interplay between the constraints and dynamical field components in gravity, 
no consistent stochastic approximation has been so far developed for theories that include 
dynamical gravity. This is a pity, since this leaves us with an incomplete understanding 
of the dynamics of quantum fields on de Sitter space. 
The task is further complicated by
the fact that only a few perturbative results are known that include quantum 
gravity~\cite{Tsamis:1996qk,Tsamis:1996qq,Tsamis:1996qm,Miao:2005am,Miao:2006gj,Miao:2012bj,
Janssen:2008dw,Kahya:2007bc,Park:2011kg,Park:2011ww}. 
The situation is much better as regards perturbative results for various field theories
on de Sitter~\cite{Onemli:2002hr,Onemli:2004mb,Brunier:2004sb,Janssen:2009nz,Janssen:2008px,
Koksma:2009tc,Prokopec:2002jn,Prokopec:2002uw,Prokopec:2003iu,Kahya:2009sz}. Albeit useful, 
many of these results cannot be trusted at late times, when perturbative treatment breaks down.
Therefore, one of the most burning unanswered questions of the physics of de Sitter space is: 
\begin{itemize}
\item[]
{\it What is the effective field theory that governs dynamics and probability distribution of  
the infrared fields in interacting field theories on de Sitter space?} 
\end{itemize}
That question is particularly interesting in the context of
eternal inflation, as there one expects that any time dependence can be viewed as a dependence on
some physical scale. 

 In order to make progress towards such an effective theory of inflation, note that
 -- in the spirit of the renormalization group approach to effective field theories --  
the time dependence in~(\ref{Fokker-Planck}) can be viewed as a (physical) scale dependence,
where $\mu = \mu_0 {\rm e}^{-Ht}$ (this dependence originates from the exponential expansion 
of physical scales in de Sitter space on super-Hubble length scales), such that 
$\partial_t \rightarrow -H\mu\partial_\mu$. With this observations Eq.~(\ref{Fokker-Planck}) can be  
rewritten as a renormalization group (RG) equation,
\begin{equation}
 \mu\partial_\mu \rho + \frac{1}{3H^2}\partial_\phi\left(V^\prime \rho\right) + \frac{H^2}{8\pi^2}\partial_\phi^2\rho = 0
\,.
 \label{Fokker-Planck:RG}
 \end{equation}
This heuristic equation determines the PDF $\rho=\rho(\phi(\vec x),\mu)$ 
as a function of the field $\phi$ and the scale $\mu$, which should be interpreted 
as the physical scale on which the field varies. 
An important question is how the PDF changes if one integrates out fluctuations above
the scale $\mu$. In the spirit of RG, one gets the effective PDF
$\rho_{\rm eff}(\phi,\mu)\equiv \rho_\mu(\phi)$ when 
fluctuations above that scale are integrated out. We shall postpone a concrete calculation of
$\rho_\mu(\phi)\propto {\rm exp}\Big(-\frac{8\pi^2}{3H^4}V_\mu(\phi)\Big)$ 
for future work, and concentrate here on understanding
the behavior of $\rho_\mu(\phi)$ and $V_\mu(\phi)$ in the limit when $\mu\rightarrow 0$.
This PDF we refer to as the effective PDF, $\rho_{\rm eff}$, and it signifies the 
probability distribution of (deep infrared) fields on the asymptotic timelike future infinity 
of de Sitter space.
\begin{figure}[ht]
\includegraphics[scale=0.86]{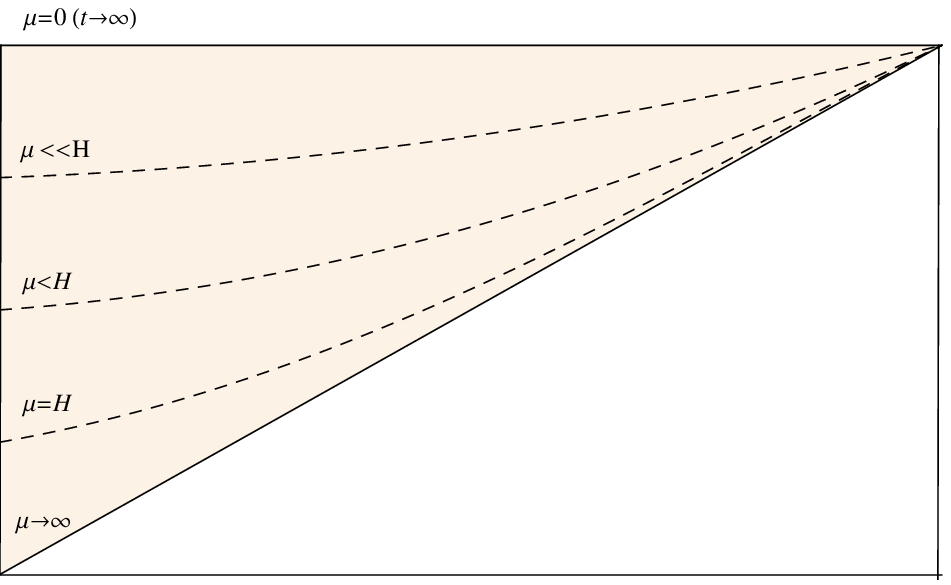}
\hskip 0.8cm
\includegraphics[scale=0.98]{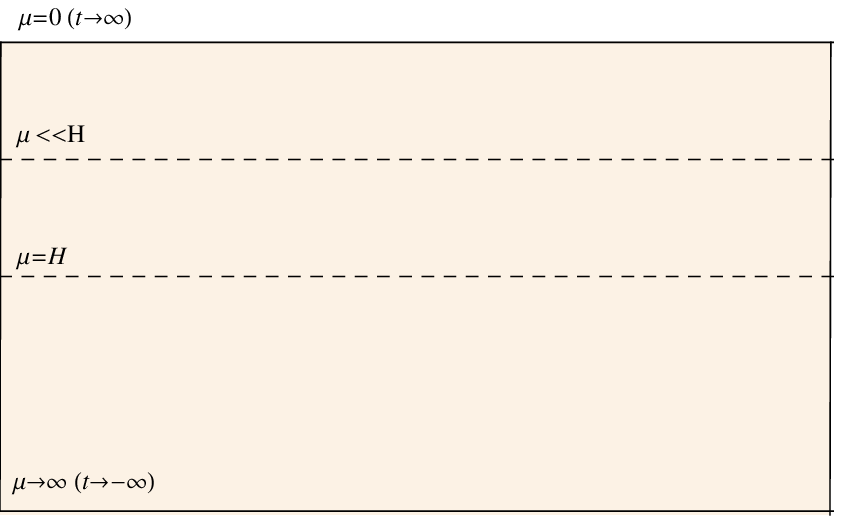}
\caption{ 
\small The Carter-Penrose diagram of de Sitter space in flat coordinates used in
this paper (left panel) and in global coordinates with positively curved spatial sections
(right panel). The curves corresponding to constant physical scales $\mu\rightarrow\infty$, 
$\mu=H$, $\mu<H$, $\mu\ll H$ and $\mu=0$ are also shown.
Even though flat coordinates cover only 1/2 of de Sitter space (shaded), 
asymptotically the surface $\mu\rightarrow 0$ 
(or equivalently, the future timelike infinity surface $i^+$, on which $t\rightarrow \infty$) 
is equal in both coordinates.
We hence expect -- and claim -- that the effective theory that governs the field distribution 
on the surface $\mu=0$ is independent on coordinates. 
\label{fig:CarterPenrose-dS}
 }
\end{figure}
The program we propose here is illustrated on the Carter-Penrose diagrams in 
figure~\ref{fig:CarterPenrose-dS}. 
The effective field theory (the effective potential and the corresponding
effective probability distribution function) we discuss in this paper 
lives on the $\mu=0$ ($t=\infty$) surface, and it is independent on the coordinates one uses.
One can use this  effective field theory to calculate the correlators of the fields that 
originate from (correlated) sub-Hubble vacuum fluctuations at some early time, and 
correspond to very large scale correlators of deep infrared fields at
the asymptotic timelike future infinity of de Sitter. 
Albeit Eq.~(\ref{Fokker-Planck:RG}) is appealingly simple, 
it is not obtained by a rigorous derivation, and we shall address
elsewhere the problem how to formally construct an effective field theory on de Sitter
valid on some finite physical scale.

 The effective theory approach advocated here becomes even better motivated when
one recalls the observed equivalence between the zero mode partition function $Z_E(\phi_{0E})$ 
on Euclidean de Sitter space 
and the PDF in stochastic approach at asymptotically late times 
$\rho(\phi,t\rightarrow\infty)$~\cite{Rajaraman:2010xd},
\begin{equation}
  Z_E(\phi_{0E}) = \rho(\phi,t\rightarrow\infty)
  \,,
\label{equivalence of Euclidean and stochastic part fn}
\end{equation}
where $Z_E$ denotes the Euclidean partition function, $\phi_{0E}$ is the zero field mode
on Euclidean de Sitter space.~\footnote{Of course, this equivalence holds only when 
the $t\rightarrow \infty$ limit of $\rho(\phi,t)$ exists, which is not guaranteed.}
 Namely, while the Euclidean space zero mode is distributed
according to $Z_E[\phi_{0E}]$, the corresponding effective action,
$V_{\rm E eff}$, 
can be constructed by making use of a  Legendre transformation. 
For a recent perturbative study of the Euclidean zero mode correlators see 
Ref.~\cite{Beneke:2012kn}.
Inspired by this observation, an adaptation of this approach to
stochastic inflation is the approach we advocate in this paper.

\medskip

 In order to make our study specific, we shall consider an $O(N)$ symmetric scalar field theory,
 whose tree level action is of the form, 
 %
\begin{equation}
S = \int d^D x \sqrt{-g} 
 \Bigg[-\frac12 g^{\mu\nu}\sum_{a=1}^N (\partial_\mu\phi_a)(\partial_\nu \phi_a)
  -\frac 12 m_0^2\sum_{a=1}^N\phi_a^2 - \frac{\lambda_0}{4N}\bigg(\sum_{a=1}^N\phi_a^2\bigg)^2\,
 \Bigg]
\,.
\label{O(N) action}
\end{equation}
%
where $m_0$ and $\lambda_0$ denote a bare field's mass and a bare quartic self-coupling, respectively,
$g_{\mu\nu}$ is the metric tensor, $g^{\mu\nu}$ its inverse and
$g={\rm det}[g_{\mu\nu}]$. The metric signature we use is $(-,+,+,..)$.
Upon a standard renormalization procedure~\cite{Prokopec:2011ms}, $m_0^2\rightarrow m^2 \equiv -\mu^2$,
and  $\lambda_0\rightarrow \lambda>0$, where now $m^2$ and $\lambda$ are finite
renormalized parameters. The resulting action can be used as a starting point for
stochastic theory. 
When $m^2<0 $ ($\mu^2>0$) the theory exhibits symmetry
breaking in Minkowski space, which is what we assume throughout this work. 
In the case when $N=1$, one obtains the action of a real scalar field, 
and we shall assume that the (renormalized) action is of the form, 
\begin{equation}
S[\phi] = \int d^D x \sqrt{-g} \left[-\frac12 g^{\mu\nu} (\partial_\mu\phi)(\partial_\nu \phi)
          -\frac 12 m^2 \phi^2 - \frac{\lambda}{4!}\phi^4\right]
\,.
\label{O(1) action}
\end{equation}
When studied on Minkowski space, the vacuum of the theory~(\ref{O(N) action}) exhibits a scalar field condensate
that breaks the $O(N)$ symmetry down to $O(N\!-\!1)$ (in the case when $N=1$, 
$O(0)$ means that the symmetry $O(1)\equiv Z_2$ is completely broken by the condensate). 
This spontaneous symmetry breaking (SSB) results in a non-trivial vacuum ${\cal M}=O(N)/O(N\!-\!1)$.
Excitations along these vacuum directions are massless, 
and they are known as Goldstone bosons. For example, when 
$N=3$, the vacuum corresponds to the two dimensional sphere, ${\cal M}=O(3)/O(2)\sim S^2$, such that 
there are two massless Goldstone bosons, corresponding to excitations along 
the two orthogonal directions on $S^2$. The vacuum of the theory is said to be trivial
if $\vec \phi(\vec x)$ maps all of the physical space $R^3$ into a point on field space, 
$\|\vec \phi\|=\phi_0 =\mu\sqrt{N/\lambda}$ (here $N=3$). If however $\vec \phi(\vec x)$ maps 
the 2-sphere $\|\vec x\|\rightarrow \infty$ of the physical space onto a 2-sphere of the internal space, 
the vacuum is said to be topologically nontrivial 
(the second homotopy group of the vacuum manifold is non-trivial, 
$\pi_2({\cal M})=Z$~\cite{Kibble:1980mv}).
This configuration is known as the global monopole, and once formed it is (topologically) stable.
Global monopoles are an example of topological defects, which have been studied in the 1980s and 1990s
as an alternative to inflation that can seed Universe's structure formation. 
When confronted with modern cosmic microwave background observations~\cite{Ade:2013xla},
these theories had to be abandoned. However, global defects have been invoked to drive 
inflation~\cite{Vilenkin:1994pv} 
or provide a possible explanation for dark energy~\cite{BuenoSanchez:2011wr}.

 The main purpose of this paper is to prove that no topological defects can survive 
in eternal inflation. In other words, deep infrared fields cannot see 
symmetry breaking and hence cannot form a condensate.
For earlier work on this problem see Refs.~\cite{Vilenkin:1982wt,Ford:1985qh,Ratra:1984yq}.
The infrared effects from abundant particle production 
(in arbitrary number of space-time dimensions) 
in eternal inflation are so strong that they will eventually restore symmetry of any tree level 
potential of the form~(\ref{O(N) action}--\ref{O(1) action}) 
and destroy any defects that might have formed on 
(sub-)Hubble scales~\cite{Basu:1991ig,Basu:1992ue,Basu:1993rf}. 
As an important side remark, we note that our results 
suggest that very deep infrared modes will exhibit
an enhanced non-Gaussianity of a non-perturbative character. 

 The paper is organized as follows. In section~\ref{Symmetry restoration} we prove by explicit calculation
that no condensate can form on de Sitter space in a theory~(\ref{O(N) action})
which at tree level exhibits a symmetry breaking ($m^2<0$). For pedagogical reasons, we present
a separate proof for $O(1)$, $O(2)$ and the general $O(N)$ case.
In section~\ref{Effective potential} we make use of the Legendre transform and show 
that the effective field theory at zero physical scale is convex for an arbitrary value of the background field.
Again, we firstly study the $O(1)$ case, and then we prove convexity in the general $O(N)$ 
case. We also derive analytical approximations of the effective potential
for small and large background field values. 
Finally, in section~\ref{Discussion} we summarize our main results, and point at future research directions.


\section{Symmetry restoration}
\label{Symmetry restoration}

 In order to prove that the symmetry in an $O(N)$ model is restored, it is sufficient to show that 
the curvature of the effective potential governing deep infrared field fluctuations 
is positive at the origin. We shall prove this by {\it reductio ad absurdum}. Namely, we 
shall show that the assumption that a condensate forms leads to a contradiction. We first consider 
a real scalar field, and then a complex (two component) scalar field and finally 
we discuss the general $O(N)$ symmetric case.

 \subsection{A real scalar field}
 \label{A real scalar field}

 Let us now consider a real scalar field, whose action is given in~(\ref{O(1) action}).
We shall assume that the mass parameter $\mu^2=-m^2>0$, such that the theory exhibits 
a spontaneous symmetry breaking in Minkowski space, and the field develops 
a condensate $\phi=\pm\phi_0=\pm\mu\sqrt{6/\lambda}$ in its trivial vacuum state. 
In addition there are (topologically stable) domain wall solutions 
in which the field asymptotically condenses to
$\phi=\pm\phi_0$ for $z\rightarrow \pm \infty$ (where $z$ is some spatial direction). 
Domain walls could have formed in the early Universe
by the Kibble mechanism~\cite{Kibble:1980mv}, if at high temperatures the symmetry was restored by
thermal field fluctuations. In this case, as the Universe expands it cools down and the model undergoes 
a phase transition.

 We shall now consider the model~(\ref{O(1) action}) in inflationary (de Sitter) background.
We shall argue that, while short scale field fluctuations can see a symmetry breaking potential, 
very long scale fluctuations see necessarily a potential in which the symmetry is restored.
Let us begin our analysis by varying the action~(\ref{O(1) action}) with respect to $\phi$.
On super-Hubble the derivative terms can be dropped and the equation of motion 
for the mean (background) field $\phi_{\rm b}$ simplifies to,
\begin{equation}
\left[-\mu^2+\frac{\lambda}{2}\langle\delta\phi^2\rangle_{\rm fin}+\frac{\lambda}{6}\phi_{\rm b}^2\right]
   \phi_{\rm b} = 0
\,,
\label{stationary point:1}
\end{equation}
where the second term includes both the contribution from the fluctuations $\delta\phi=\phi-\phi_{\rm b}$,
that can be estimated by solving
the Fokker-Planck equation~(\ref{Fokker-Planck}) for stochastic inflation
(while the sub-Hubble fluctuations contribute mainly to renormalize the coupling parameters, 
the super-Hubble field fluctuations are captured in $\langle\delta\phi^2\rangle_{\rm fin}$ 
and can be estimated by stochastic inflation,  
$\langle\delta\phi^2\rangle_{\rm fin}=\langle\delta\phi^2\rangle_{\rm stoch}$).

Ignoring the $\phi_{\rm b}=0$ solution (which is a local maximum), Eq.~(\ref{stationary point:1})
implies,
\begin{equation}
\phi_{\rm b}^2=\phi_0^2-3\langle\delta\phi^2\rangle_{\rm stoch} \geq 0
\,.
\label{stationary point:2}
\end{equation}
Assuming a stationary probability distribution $\rho(\phi,t)=\rho(\phi)$, Eq.~(\ref{Fokker-Planck}) 
simplifies, and it is easily solved by 
\begin{eqnarray}
\rho_{1}(\phi) &=& \frac{1}{Z_1}\exp\Big(-\frac{8\pi^2}{3H^4}V_1\Big)
\,;\qquad 
 Z_1(\zeta) =   \dfrac{\pi}{2}\phi_0{\rm e}^{\zeta /2} 
       \Big[I_{1/4}(\zeta/2) + I_{-1/4}(\zeta/2)\Big]
\nonumber\\
 V_1 &=& -\frac{\mu^2}{2!}(\phi^2) + \frac{\lambda}{4!}(\phi^2)^2 
  \;\Longrightarrow\;  \frac{8\pi^2}{3H^4}V_1  = \zeta\Big(\frac{\phi}{\phi_0}\Big)^2
           \bigg[\Big(\frac{\phi}{\phi_0}\Big)^2-2\bigg]
 \,, 
\label{rho V1}
\end{eqnarray}
where $\zeta=(4\pi^2\mu^4)/(\lambda H^4)$ denotes a dimensionless `inverse coupling' parameter. 
The second way of writing $V_1$ in~(\ref{rho V1}) is suggestive, as it indicates that
-- when written as a function of the rescaled field $\phi/\phi_0$  
-- all properties of the stationary $\rho$ in~(\ref{rho V2})
can be parametrized in terms of just one parameter: 
the inverse coupling parameter $\zeta$. 
Thus in the limit when $\zeta\rightarrow \infty$ the theory is weakly coupled, while 
in the limit when $\zeta\rightarrow 0$ the theory becomes strongly coupled. 
While solution~(\ref{rho V1}) has been known for quite a while~\cite{Starobinsky:1994bd}, 
we give it a slightly different physical interpretation.
For us Eq.~(\ref{rho V1}) represents the PDF for the (stationary) modes 
that vary over approximately the Hubble scale.
Albeit the PDFs that exhibit a nontrivial time dependence are interesting of their own right, 
for simplicity we do not study them here. Namely, Starobinsky and
Yokoyama~\cite{Starobinsky:1994bd}
have shown that, after a sufficient amount of time, every initial state necessarily reduces 
to the PDF in~(\ref{rho V1}), making the stationary distribution~(\ref{rho V1}) 
an attractor. 

\begin{figure}[t!]
\includegraphics[scale=1.2]{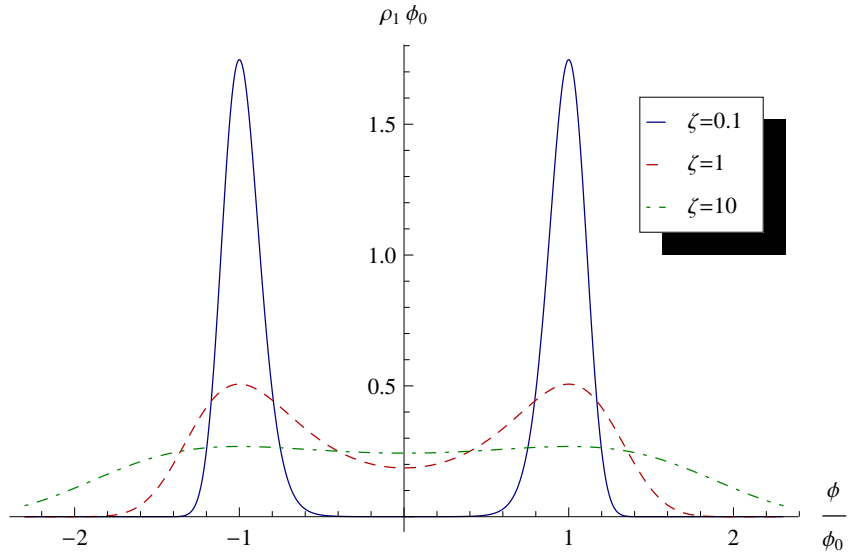}
 \caption{\small The PDF $\phi_0\times\rho_1(\phi)$ as a function of $\phi/\phi_0$ 
for the Hubble scale modes. 
The three curves represent $\zeta = 0.1$  (solid blue double peaked curve),
$\zeta = 1$ (dashed red middle curve) and $\zeta = 10$ (dot-dashed green flattest curve), 
where $\zeta=4\pi^2\mu^4/(\lambda H^4)$ is the dimensionless parameter characterizing 
the inverse coupling strength.
\label{fig:initial stochastic PDF}
}
\end{figure}
In figure~\ref{fig:initial stochastic PDF}
we show the initial PDF for these stationary (mainly) Hubble scale modes. 
We shall refer it to as the PDF for Hubble scale modes,
and the corresponding $V=V(\phi,\mu\sim H)$ is the effective (Hubble scale) potential. 
Of course, this potential still exhibits symmetry
breaking, which can be seen from the characteristic double peak structure of the PDF
in figure~\ref{fig:initial stochastic PDF}, which is  
more pronounced for strong couplings ($\zeta\ll 1$). 
Nevertheless, as we show below, the inclusion of these fluctuations
into~(\ref{stationary point:2}) is sufficient to prevent background field condensation. 
Indeed, inserting 
\begin{equation}
\langle\delta\phi^2\rangle_{\rm stoch}= \langle \phi^{\,2}\rangle_{\rm stoch}+\phi_{\rm b}^2 
\nonumber
\end{equation}
into Eq.~(\ref{stationary point:2}) results in
\begin{eqnarray}
\phi_{\rm b}^{2}
 &=&\frac{\phi_0^2}{4}
   \left(1-3\langle {\phi^2}/{\phi_0^2}\rangle_{\rm\! stoch} \right)\geq 0
\label{expec value O(1):condition}
\\
 \bigg\langle \frac{\phi^2}{\phi_0^2}\bigg\rangle_{\rm\! stoch} 
 &=& \dfrac{1}{2} \left( 1+ \frac{I_{3/4}(\zeta/2)+I_{-3/4}(\zeta/2)}
                                 {I_{1/4}(\zeta/2)+I_{-1/4}(\zeta/2)}
\,\right),
\label{expec value O(1)}
\end{eqnarray}
where 
\begin{equation}
 I_\nu(z)= \sum_{k=0}^{\infty} \frac{1}{\Gamma(k + \nu + 1) k!} \left(\frac{z}{2}\right)^{2 k + \nu}
\label{Bessel I}
\end{equation} 
is the series representation around $z=0$ for the modified Bessel's function of the first kind.
When~(\ref{expec value O(1)}) is inserted into Eq.~(\ref{expec value O(1):condition}) one gets, 
\begin{equation}
\phi_{\rm b}^{2}=-\frac{\phi_0^2}{8}
\left(1+3 \frac{I_{-3/4}(\zeta/2)+I_{3/4}(\zeta/2)}{I_{-1/4}(\zeta/2)+I_{1/4}(\zeta/2)}\right)\geq 0
\,,
\label{expec value O(1):b}
\end{equation}
which is impossible to satisfy, thus completing the proof.
The proof is an immediate consequence of the observation that, if $\nu>-1$, $I_\nu(z)$ is strictly positive 
for all $z>0$. This is true because for $\nu>-1$ all coefficients in the series~(\ref{Bessel I}) 
for $I_\nu(z)$ are positive when $\nu>-1$, and this condition is satisfied for
all Bessel's functions occurring in~(\ref{expec value O(1):b}).
We have thus proved that, when the field fluctuations evaluated within stochastic theory of inflation 
are taken account of, no phase transition in the $O(1)$ model~(\ref{O(1) action}) can occur,
 {\it i.e.} there can be no field condensate.

\subsection{The $O(N)$ model}
\label{The O(N) model}

 The proof in the $O(N)$ symmetric model~(\ref{O(N) action})
is analogous to the proof in the $O(1)$ case. 
The main complication is the enlarged symmetry, 
and we shall devote some attention to explain the resulting differences. 
In the $O(N)$ case the field $\vec \phi=(\phi_a)$ ($a=1,..,N$) is a $N$-component vector, and  
when a mean field condensate is present, a suitable orthogonal rotation can bring the condensate 
direction into the direction of $\phi_1$, such that $\phi_{\rm b} = \phi_{1\rm b}$ and 
$(\phi_0)^2 = (\phi_1)_0^2=N\mu^2/\lambda$. The  generalization of~(\ref{stationary point:1})    
to the $O(N)$ case is ({\it cf.} Ref.~\cite{Prokopec:2011ms})
%
\begin{equation}
\phi_{\rm b}^2=(\phi_1)^2=\phi_0^2-3\langle(\delta\phi_1)^2\rangle_{\rm stoch}
    -(N\!-\!1)\langle(\delta\phi_2)^2\rangle_{\rm stoch}\geq 0
\,,
\label{O(N):symmetry breaking:condensate}
\end{equation}
where
%
\begin{eqnarray}
\langle (\delta\phi_1)^2 \rangle &=& \langle (\phi_1-\phi_{\rm b})^2\rangle
      = (\phi_{\rm b})^2 + \langle (\phi_1)^2\rangle 
      = (\phi_{\rm b})^2 + \frac{1}{N}\langle (\vec \phi\,)^{\,2}\rangle
\label{Delta11 in terms of Stoch}\\
\langle (\delta\phi_2)^2 \rangle  &=& \langle (\phi_2)^2\rangle
      =  \frac{1}{N}\langle (\vec \phi\,)^{\,2}\rangle
\,.
\label{Delta22 in terms of Stoch}
\end{eqnarray}
%
Because the scalar potential in~(\ref{O(N) action})
depends only on $\sum_{a=1}^N(\phi_a)^2=(\vec \phi\,)^2$,
the linear term in~(\ref{Delta11 in terms of Stoch}) does not contribute,
and $\langle (\delta\phi_i)^2\rangle = \langle (\delta\phi_j)^2\rangle$ ($\forall i,j\in\{1,..,N\}$).
Inserting Eqs.~(\ref{Delta11 in terms of Stoch}--\ref{Delta22 in terms of Stoch})
into~(\ref{O(N):symmetry breaking:condensate}) results in
\begin{equation}
\phi_{b}^2=\frac{\phi_0^2}{4}
   \left(1-\frac{N+2}{N}\langle (\vec \phi/\phi_0)^{\,2}\rangle_{\rm stoch}\right)\geq 0
\,.
\label{O(N):symmetry breaking:condensate:2}
\end{equation}
We shall now show that this inequality has no non-trivial solution,  
\textit{i.e.} that no condensate can form for stationary solutions in stochastic inflation.

 But, before we consider the general case, it is instructive to consider 
the $O(2)$ model (complex scalar field), since in this case the results are particularly simple. 
We have
\begin{eqnarray}
\rho(\phi_1,\phi_2) &=& \frac{1}{Z_2}\exp\Big(-\frac{8\pi^2}{3H^4}V_2\Big)
\,,\qquad  
 \frac{1}{Z_2} = \frac{1}{\phi_0^2}\frac{2}{\pi^{3/2}}
     \frac{\sqrt{\zeta}{\rm e}^{-\zeta}}{1+{\rm erf}\big(\sqrt{\zeta}\big)}
     \,,\qquad \zeta = \frac{4\pi^2\mu^4}{3\lambda H^4}
\nonumber\\
 V_2 &=& -\frac{\mu^2}{2}(\phi_1^2+\phi_2^2) + \frac{\lambda}{8}(\phi_1^2+\phi_2^2)^2
\;\Rightarrow \; \frac{8\pi^2}{3H^4}V_2 = \zeta(\vec \phi/\phi_0)^2\big[(\vec \phi/\phi_0)^2-2\big]
\,,
\label{rho V2}
\end{eqnarray}
where ${\rm erf}(z)=(2/\sqrt{\pi})\int_0^z dt{\rm e}^{-t^2}$ denotes the \textit{error} function
and we normalized $\rho$ by demanding $\int_{-\infty}^{\infty} d\phi_1d\phi_2 \rho = 1$.

In order to make progress on inequality~(\ref{O(N):symmetry breaking:condensate:2}),
we need to calculate $\langle\vec \phi^{\,2}\rangle_{\rm stoch}$. Making use
of~(\ref{rho V2}) for $N=2$ we immediately get,
\begin{eqnarray}
\langle\vec \phi^{\,2}\rangle_{\rm stoch}
  &=& \frac{\pi\phi_0^4}{Z_2}\int_0^\infty d\varphi^2 \varphi^2\exp\{-\zeta[(\varphi^2-1)^2-1]\}
  \nonumber\\
   &=& 2\phi_0^2\frac{\sqrt{\zeta/\pi}}{1+{\rm erf}(\sqrt{\zeta}\,)}\int _{-1}^\infty du(u+1){\rm e}^{-\zeta u^2}
\nonumber\\ 
   &=& \phi_0^2\left(1+\frac{{\rm e}^{-\zeta}}{\sqrt{\pi\zeta}[1+{\rm erf}(\sqrt{\zeta}\,)]}\right)  
\,,
\label{phi squared}
\end{eqnarray}
where $\varphi^2=(\phi_1^2\!+\!\phi_2^2)/\phi_0^2$ and $u=\varphi^2-1$.
Upon inserting this result into~(\ref{O(N):symmetry breaking:condensate:2}) with $N=2$ 
yields
\begin{equation}
\phi_{\rm b}^2=-\frac{\phi_0^2}{4}
  \Bigg[1 + \frac{2{\rm e}^{-\zeta}}{\sqrt{\pi\zeta}[1+{\rm erf}(\sqrt{\zeta}\,)]}\Bigg] > 0
\,.
\label{O(N):symmetry breaking:condensate:3}
\end{equation}
 Now, since ${\rm erf}(z)>0$ ($\forall z>0$) Eq.~(\ref{O(N):symmetry breaking:condensate:3})
 cannot be satisfied for any $\zeta> 0$, proving the impossibility of condensate formation 
for the $O(2)$ case. 

\bigskip

 We are now ready to consider symmetry breaking in the general $O(N)$ symmetric case.
In this case the PDF for a stationary state is the following generalization 
of Eqs.~(\ref{rho V2}) 

\begin{eqnarray}
\rho &=& \frac{1}{Z}\exp\Big(-\frac{8\pi^2}{3H^4}V\Big)
\,,\qquad   \qquad  \zeta = \frac{2N\pi^2\mu^4}{3\lambda H^4}
 \,,\qquad \phi_0^2 = \frac{N\mu^2}{\lambda}, 
\nonumber\\
Z &=& \phi_0^N \frac{\Omega(S^{N\!-\!1})}{4\zeta^{N/4}}  
   \left[\Gamma\Big(\frac{N}{4}\Big)\times{}_1F_1\Big(\frac{N}{4};\frac{1}{2};\zeta\Big)
            + 2\sqrt{\zeta} \Gamma\Big(\frac{N\!+\!2}{4}\Big)  
     \times{}_1F_1\Big(\frac{N\!+\!2}{4};\frac{3}{2};\zeta\Big)\right]
\nonumber\\
 \, V &=& -\frac{\mu^2}{2}\bigg(\sum_{a=1}^N\phi_a^2\bigg) 
             + \frac{\lambda}{4N}\bigg(\sum_{a=1}^N\phi_a^2\bigg)^2
\;\Rightarrow \; \frac{8\pi^2}{3H^4}V = \zeta\bigg(\frac{\vec \phi}{\phi_0}\bigg)^2
    \bigg[\bigg(\frac{\vec \phi}{\phi_0}\bigg)^2-2\bigg]
\,,
\label{rho VN}
\end{eqnarray}
where $\Omega(S^{N\!-\!1})=2 \pi^{N/2}/\Gamma(N/2)$ denotes the volume (surface area) of 
the unit $N\!-\!1$ dimensional sphere.
The two point function is then:
\begin{eqnarray}
 \langle\vec \phi^{\,2}\rangle_{\rm stoch} 
  &=& \Omega(S^{N\!-\!1})\frac{\phi_0^{N+2}}{Z} \int^{\infty}_{0}d\varphi \varphi^{N+1}
   {\rm e}^{-\zeta[(\varphi^2-1)^2-1]}
\nonumber\\
  &=& \Omega(S^{N\!-\!1})\frac{\phi_0^{N+2}}{Z} \int^{\infty}_{-1}du(u+1)^{N/2}
   {\rm e}^{-\zeta u^2}{\rm e}^\zeta
\nonumber\\
  &=&\frac{\phi_0^2}{\sqrt{\zeta}}
 \times \frac{\Gamma\left(\frac{N\!+\!2}{4}\right)  
                  \times {}_1F_1\left(\frac{N\!+\!2}{4};\frac{1}{2};\zeta\right)
              + 2\sqrt{\zeta}\Gamma\left(\frac{N}{4}\!+\!1\right) 
                  \times{}_1F_1\left(\frac{N}{4}\!+\!1;\frac{3}{2};\zeta\right)}
            {\Gamma\left(\frac{N}{4}\right) 
               \times {}_1F_1\left(\frac{N}{4};\frac{1}{2};\zeta\right)
             + 2\sqrt{\zeta}\Gamma\left(\frac{N\!+\!2}{4}\right) 
                   \times{}_1F_1\left(\frac{N\!+\!2}{4};\frac{3}{2};\zeta\right)} 
\,,
\label{phi2 - O(N)}
\end{eqnarray}
where $\varphi^2=\vec \phi^{\;2}/\phi_0^{\,2}$ 
and $u=\varphi^2-1$.
Upon inserting~(\ref{phi2 - O(N)}) into~(\ref{O(N):symmetry breaking:condensate:2}) we get
\begin{equation}
 \phi_{\rm b}^2 = \frac{\phi_0^2}{4}F(\zeta, N)
\label{phi b: N case}
\end{equation}
where 
%
\begin{eqnarray}
 F(\zeta, N)=1-\frac{N\!+\!2}{N}\frac{2\Gamma\left(\frac{N}{4}\!+\!1\right) 
           \times{}_1F_1\left(\frac{N}{4}\!+\!1;\frac{3}{2};\zeta\right)
    +\frac{1}{\sqrt{\zeta}}\Gamma\left(\frac{N\!+\!2}{4}\right)  
        \times{}_1F_1\left(\frac{N\!+\!2}{4};\frac{1}{2};\zeta\right)}
   {\Gamma\left(\frac{N}{4}\right)
           \times{}_1F_1\left(\frac{N}{4};\frac{1}{2};\zeta\right)
    + 2\sqrt{\zeta}\Gamma\left(\frac{N\!+\!2}{4}\right) 
              \times{}_1F_1\left(\frac{N\!+\!2}{4};\frac{3}{2};\zeta\right)}
\,.
\label{O(N):symmetry breaking:condensate:4}
\end{eqnarray}
%
In the appendix we show that $F(\zeta, N)<0$ for any integer $N>0$ and $\zeta>0$,
which completes the proof that there can be no mean field condensate in an $O(N)$ symmetric
scalar field theory in de Sitter space.



\section{Effective potential}
\label{Effective potential}

 While in section~\ref{Symmetry restoration} we show that inflationary fluctuations
treated within the stochastic formalism are strong enough to prevent 
formation of a scalar condensate in any $O(N)$ symmetric scalar theory in de Sitter inflation, 
here we prove a much more powerful theorem. Namely, 
we show that the effective potential governing the PDF of 
very long wave length fluctuations
in inflation is strictly convex (for arbitrary value of the field).
This represents a general proof that deep infrared modes in de Sitter space
see a symmetry restoring potential, and hence their PDF must be peaked at zero field value.
Inspired by the observation that the partition function of the zero Euclidean mode 
equals the PDF for time independent field 
configurations~(\ref{equivalence of Euclidean and stochastic part fn})
we shall use the effective action formalism to construct 
$V_{\rm eff}$ and the corresponding $\rho_{\rm eff}$.
 We first consider the $O(1)$ case, and then subsequently the general $O(N)$ symmetric model.

\subsection{The $O(1)$ model}
\label{The O(1) model}

 Since $\rho(\phi)$ yields a PDF for $\phi$ that varies on the Hubble scale,
it is reasonable to posit that $\rho_{\rm eff}$ obtained by the conventional 
field theoretic technique of Legendre transform~\cite{Peskin:1995ev}
will give a PDF for deep infrared ($\mu\rightarrow 0$) 
scalar field fluctuations $\phi_{\rm b}$, whereby all
higher energy fluctuations have been integrated out. Since any 
field that originates from sub-Hubble scale fluctuations rapidly 
redshifts during inflation, the effective potential that we derive here 
yields the field distribution at the asymptotically late times ($t\rightarrow \infty$),
at the asymptotic future timelike infinity $i^+$ of de Sitter space, 
as illustrated in figure~\ref{fig:CarterPenrose-dS}. 

 Adding a source current $J(\vec x)$ to $V(\phi)$ one can define a partition function
(at a point $\vec x\,$) $Z(J(\vec x\,))$ as 
\begin{equation}
Z(J) \equiv {\rm e}^{-W(J)} = \int d\phi\rho_1(\phi){\rm e}^{J\phi}  
    =\langle{\rm e}^{J\phi}\rangle
\,.
\label{Z(J)}
\end{equation}
The effective potential $V_{\rm eff}$ is then given as a Legendre transform of 
$W(J)=-\ln[Z(J)]$,
\begin{equation}
  \frac{8\pi^2}{3H^4}V_{\rm eff}(\phi_{\rm b})= W(J)+J \phi_{\rm b}
\,;\qquad   \phi _{\rm b} = \frac{\partial \ln Z(J)}{\partial J}
\label{Veff:O(1)} 
\end{equation}
and the corresponding PDF is then 
\begin{equation}
\rho_{\rm eff}(\phi_{\rm b})
   =\frac{1}{Z_{\rm eff}}{\rm exp}\left(-\frac{8\pi^2}{3H^4}V_{\rm eff}(\phi_{\rm b})\right)  
\;;\qquad Z_{\rm eff} = \int d\phi_{\rm b}{\rm exp}\left(-\frac{8\pi^2}{3H^4}V_{\rm eff}\right)
\,.
\label{rho_eff}
\end{equation}

 In order to study convexity of $V_{\rm eff}(\phi_{\rm b})$ 
we shall make use of the equation of motion for $\phi_{\rm b}$ and its 
derivative,
 \begin{eqnarray}
 \dfrac{\partial V_{\rm eff}(\phi_{\rm b})}{\partial \phi_{\rm b} }=J 
\; ;\qquad 
\dfrac{\partial^{2} V_{\rm eff}(\phi_{\rm b})}{\partial \phi_{\rm b}^{2} }
    = \dfrac{\partial J(\phi_{\rm b})}{\partial \phi_{\rm b}}
\,.
 \label{derivatives1}
\end{eqnarray}
$V_{\rm eff}$ is convex if $\partial^{2} V_{\rm eff}/\partial \phi_{\rm b}^{2}>0$ 
for all $\phi_{\rm b}$. The inverse of this must also be positive,
  \begin{eqnarray}
\left(\dfrac{\partial^2 V_{\rm eff}(\phi_{\rm b})}{\partial \phi_{\rm b}^2} \right)^{-1}
 = \dfrac{\partial^{2}\ln[Z(J)]}{\partial J^2}
  = \dfrac{Z''(J)}{Z(J)} -\left( \dfrac{Z'(J)}{Z(J)}\right)^2 > 0
\,.
\label{convexity condition}
\end{eqnarray}
Now, from the definition of $Z(J)$ in Eq.~(\ref{Z(J)}) is follows  
\begin{equation}
 Z(J) = \sum_{n=0}^{\infty} \frac{\langle \phi^{n} \rangle J^{n}}{n!} 
      = \langle {\rm e}^{J\phi}\rangle
\,,
\label{Z(J):2}
\end{equation}
where 
\begin{equation}
\left\langle\frac{\phi^{2n}}{\phi_0^{2n}}\right\rangle_{\rm stoch}
   = \frac{\Gamma\big(\frac{n}{2}+\frac14\big)
                    \times {}_1F_1\big(\frac{n}{2}+\frac14;\frac12;\zeta\big)
          + 2\sqrt{\zeta}\,\Gamma\big(\frac{n}{2}+\frac34\big)
                     \times {}_1F_1\big(\frac{n}{2}+\frac34;\frac32;\zeta\big)}
 {\pi{\rm e}^{\zeta/2}\zeta^{\frac{n}{2}+\frac14}\big(I_{1/4}(\zeta/2)+I_{-1/4}(\zeta/2)\big)}
\,
\label{phi2n:stoch}
\end{equation}
and $\langle\phi^{2n+1}\rangle_{\rm stoch}=0$ ($n=0,1,2,..$). 
Furthermore, we have 
\begin{equation}
Z'(J) = \langle \phi e^{J\phi}\rangle = Z(J)\langle \phi\rangle_J
\;;\qquad 
  Z''(J) = \langle \phi^{2} e^{J\phi}\rangle = Z(J)\langle \phi^2\rangle_J
\,,
\label{some intermediate results for Z' and Z''}
\end{equation}
where $\langle\cdot\rangle_J$ denotes averaging
with respect to $\rho_J(\phi) = \rho(\phi){\rm e}^{J\phi}$, which is also positive.
With this in mind we can rewrite the convexity condition~(\ref{convexity condition}) as
\begin{eqnarray}
\left(\dfrac{\partial^2 V_{\rm eff}(\phi_{\rm b})}{\partial \phi_{\rm b}^2} \right)^{-1}
 = \langle\phi^2\rangle_J-\langle\phi\rangle_J^2 > 0
\,.
\label{convexity condition:2}
\end{eqnarray}
Now, since $\rho_J(\phi)$ is positive definite, we can make use of the 
Cauchy-Schwarz theorem for probability 
theory,~\footnote[46]{For any two functions \textit{f} and \textit{g} 
of stochastic variable $\phi$ distributed according to a \textit{positive} 
probability distribution function $\textit{$\rho=\rho(\phi)$}$, 
if the integral $\langle fg \rangle = \int f(\phi)g(\phi)\rho(\phi)d\phi$ exists, 
then $\langle fg \rangle$ is a well defined (commutative, bi-linear and positive-definite) 
scalar product. Then the {\it Cauchy-Schwarz theorem} holds,
according to which the following inequality must be satisfied:
%
\[
 \langle fg \rangle^{2} \leq  \langle f^2\rangle  \langle g^2 \rangle
\,.
\]
%
Equality holds only when $f$ and $g$ are linearly dependent, {\it i.e.} when 
$f\propto g$.
\label{footnote:CauchySchwarz}
}
which for $f=1$ and $g=\phi$ (see footnote~[61]) reads 
\begin{equation}
 \langle\phi\rangle_J^2 < \langle \phi^2\rangle_J
\,,
\label{Cauchy-Schwarz:2}
\end{equation}
where the strong inequality follows because  $1$ and $\phi$ 
are linearly independent functions. This immediately implies 
that the (strong) convexity
condition~(\ref{convexity condition:2}) for $V_{\rm eff}$ is satisfied, completing the proof.

\medskip

\subsection{Analytical approximations for $V_{\rm eff}$}
\label{Analytical approximations for the effective potential}

\begin{figure}[ttt]
\includegraphics[scale=0.96]{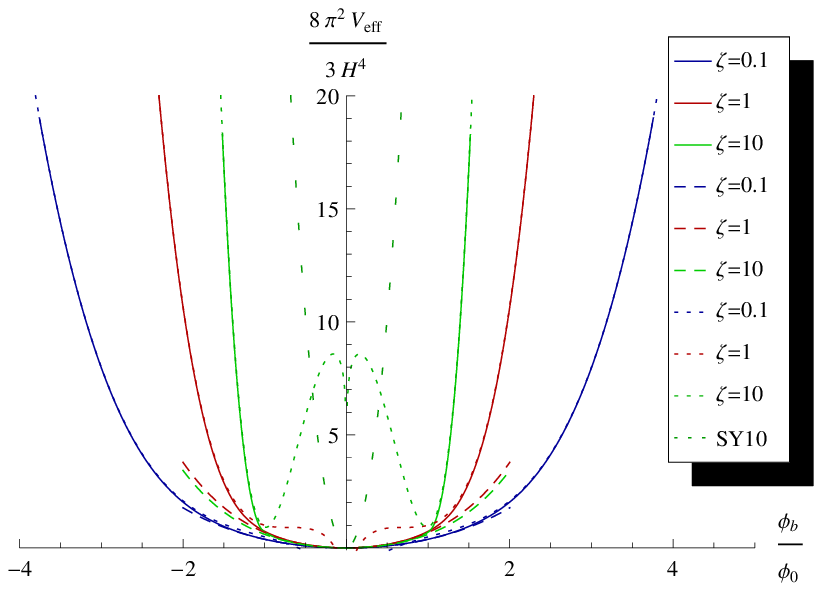}
\hskip -0.9cm
\includegraphics[scale=0.89]{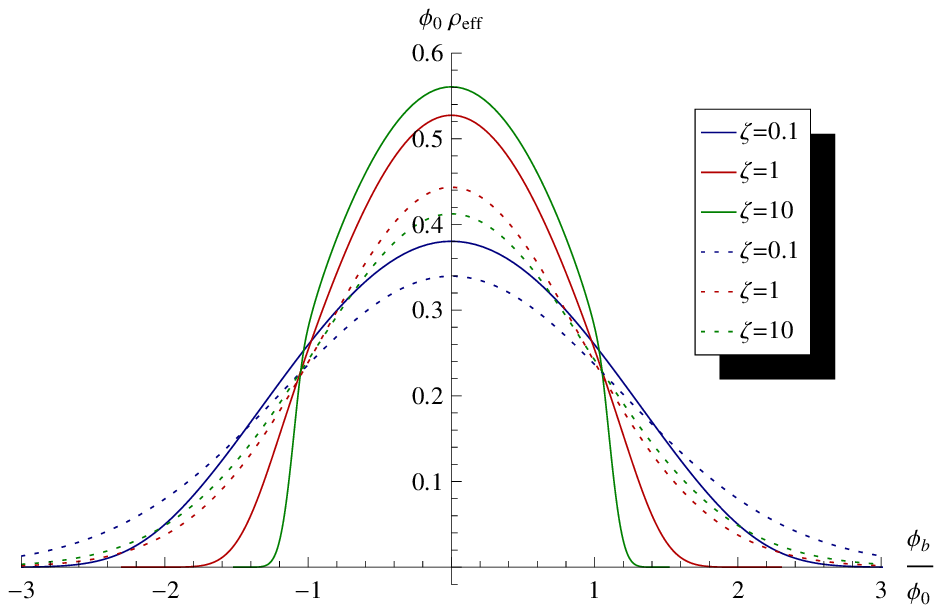}
 \caption{\small {\it Left panel:} $V_{\rm eff}$ as a function of
$\phi_{\rm b}$ for $\zeta = 0.1,1,10$ (solid blue innermost curve, 
solid red middle curve and solid green outermost curve). 
We also show the weak field analytic effective potential~(\ref{Veff:origin})
(dotted curves), the analytic expression~(\ref{Veff:asymptotic}) for
large $\phi_{\rm b}$ (dashed curves), and finally the Starobinky-Yokoyama 
effective potential~(\ref{effective potential: Starobinsky-Yokoyama}) 
(sparse green dashed innermost curve) just for $\zeta=10$ 
(in order not to overcrowd the plot).
 {\it Right panel:} $\rho_{\rm eff}$ as function of $\phi_{\rm b}$ for $\zeta = 0.1,1,10$ 
(solid curves, the same color code as on the left panel) 
and the best Gaussian fits close to the origin (dotted curves).
\label{fig:Veff-rhoeff}}
\end{figure}
 In figure~\ref{fig:Veff-rhoeff} we show numerical plots for
$V_{\rm eff}$ and $\rho_{\rm eff}$ 
as a function of the deep infrared background field $\phi_{\rm b}$,
for different values of the dimensionless inverse coupling parameter: $\zeta=0.1, 1, 10$.
The figures do not represent exact $V_{\rm eff}$, but instead they are based on taking a finite
number of terms in the sum~(\ref{Z(J):2}) involved in the definition of $Z(J)$.
However, we have checked that increasing the number of terms in the sum does not change 
the form of $V_{\rm eff}$ to such an extend to be visible on the plots. 
It is apparent that, for small values of 
$\phi_{\rm b}$, $V_{\rm eff}$ exhibits a simple quadratic dependence on $\phi_{\rm b}$,
implying an approximately Gaussian $\rho_{\rm eff}$
(that can be characterized by a positive mass term), 
as can be seen on the right panel of
figure~\ref{fig:Veff-rhoeff}. 
In the large field limit when $\phi_b\gg \phi_0$ 
the effective potential is, as expected, quartic, $[(8\pi^2/(3H^4)]V_{\rm eff}\sim\zeta\phi_b^4$.
However there is no good smooth matching between 
the quadratic and quartic behavior, and around $\phi_{\rm b}\sim \phi_0$ 
the quadratic behavior turns quickly into a quartic behavior as one would expect from 
a Maxwell construction. Indeed, forcing a quadratic plus quartic fit onto $V_{\rm eff}$ 
when $\zeta\gg 1$ results in a fit with a double well structure.
These observations then imply that weakly coupled theories 
exhibit large deviations from the behavior expected based on 
the Starobinsky-Yokoyama procedure, in the sense that 
small amplitude fluctuations of deep infrared fields are enhanced
when compared with the Starobinsky-Yokoyama result discussed in 
section~\ref{Symmetry restoration} and below.

 In fact, it is worth making a detailed comparison of the mass implied by
the effective action in figure~\ref{fig:Veff-rhoeff} and that implied by 
the Starobinsky-Yokoyama procedure~\cite{Starobinsky:1994bd,Beneke:2012kn}
which, when generalized to a finite background field $\phi_{\rm b}$, yields, 
\begin{equation}
  m_{\rm SY}^2(\phi_{\rm b}) = -\mu^2 + \frac{\lambda}{2}\langle(\phi-\phi_{\rm b})^2\rangle_{\rm stoch}
      = \mu^2\left[3\left(\left\langle\frac{\phi^2}{\phi_0^2}\right\rangle_{\rm stoch}
              \!\!+\frac{\phi_{\rm b}^2}{\phi_0^2}\right)-1\right]
\,,
\label{mass squared}
\end{equation}
where $\langle(\phi/\phi_0)^2\rangle_{\rm stoch}$ is given in Eq.~(\ref{expec value O(1)}).
%
%
On the other hand, the (numerical) mass that measures the curvature of the effective potential 
in figure~\ref{fig:Veff-rhoeff} is given by
\begin{equation}
 M^2(\phi_{\rm b}) 
     = \frac{\partial^2\Big[\frac{8\pi^2}{3H^4}V_{\rm eff}(\phi_{\rm b})\Big]}
            {\partial(\phi_{\rm b}/\phi_0)^2}
\,.
\label{numerical mass}
\end{equation}
Now, one can easily show that expressions~(\ref{expec value O(1)}) 
and~(\ref{numerical mass}) are related as, 
\begin{equation}
 m^2(\phi_{\rm b}) = \frac{\mu^2}{4\zeta} M^2(\phi_{\rm b})
\,,
\label{relation mass vs numerical mass}
\end{equation}
which -- if the Starobinsky-Yokoyama prescription is correct --
implies the following mass term,
\begin{equation} 
 M^2_{\rm SY}(\phi_{\rm b}) =
  12 \zeta\left[\left\langle\frac{\phi^2}{\phi_0^2}\right\rangle_{\rm stoch}
              \!\!+\frac{\phi_{\rm b}^2}{\phi_0^2}-\frac13\right]
\,.
\label{relation mass vs numerical mass:2}
\end{equation}
The corresponding effective potential is then
\begin{equation} 
 \frac{8\pi^2}{3H^4}V_{\rm eff\, SY}(\phi_{\rm b}) =
  6\zeta\left[\left\langle\frac{\phi^2}{\phi_0^2}\right\rangle_{\rm\! stoch}
              \!\!-\frac13\right]\frac{\phi^2}{\phi_0^2}
      + \zeta\left(\frac{\phi_{\rm b}}{\phi_0}\right)^4
      + \frac{8\pi^2}{3H^4}V_0
\,,
\label{effective potential: Starobinsky-Yokoyama}
\end{equation}
where $V_0$ is an integration constant (a linear term in $\phi_{\rm b}$ 
is forbidden by symmetry).

 On the other hand, one can calculate the effective potential around
$\phi_{\rm b}=0$ by making use of Eqs.~(\ref{Z(J)}--\ref{Veff:O(1)}). 
Upon expanding ${\rm e}^{J\phi}$ in ~(\ref{Z(J)}) in powers of $J\phi$ 
and keeping the terms up to $(J\phi)^4$) we get  
\begin{equation}
 \frac{8\pi^2}{3H^4}V_{\rm eff} = \frac12\frac{\phi_b^2}{\langle\phi^2\rangle_{\rm stoch}}
           +\frac18\left(1-\frac13\frac{\langle\phi^4\rangle_{\rm stoch}}
                                     {\langle\phi^2\rangle_{\rm stoch}^2}\right)
             \frac{\phi_b^4}{\langle\phi^2\rangle^2_{\rm stoch}}
           +{\cal O}\Big((\phi_{\rm b}/\phi_0)^6\Big)
,
\label{Veff:origin}
\end{equation}
which yields for the mass around $\phi_{\rm b}=0$,
\begin{equation}
  M^2(0) = \frac{\phi_0^2}{\langle\phi^2\rangle_{\rm stoch}}
\,,
\label{mass at origin}
\end{equation} 
implying the following physical mass~(\ref{relation mass vs numerical mass})
at $\phi_{\rm b}=0$
\begin{equation}
  m^2(0) = \frac{3H^4}{8\pi^2\langle\phi^2\rangle_{\rm stoch}}
\,.
\label{physical mass around origin}
\end{equation}

 There is a very intriguing connection of the result~(\ref{physical mass around origin}),
 with the mean field (one loop)
result for the mass~\cite{Prokopec:2011ms}.
Recall that in $D=4$, for a field of the mass parameter $m^2$,
the mean field treatment gives $(\lambda/2)\imath\Delta(x;x)$,
where the coincident propagator is, 
$\imath\Delta(x;x)=(3H^4)/(8\pi^2m^2)$.
Now, assume that the mass $m$ is entirely created
by stochastic fluctuations, thus 
$m^2\rightarrow(\lambda/2)\langle\phi^2\rangle_{\rm stoch}$,
the one-loop mean field formula (with a stochastic flavor) 
then gives $m^2(0)=(3H^4)/(8\pi^2 \langle\phi^2\rangle_{\rm stoch})$,
which is precisely the result~(\ref{physical mass around origin}).
Of course, this is a very heuristic `derivation', and 
it is very difficult to imagine that one could arrive at that 
result by any other but the rigorous method advocated in this work.

At a first sight the 
mass~(\ref{mass at origin}--\ref{physical mass around origin})
may appear bizarre, as it exhibits a very different dependence on
the coupling parameter $\zeta\propto 1/\lambda$ from that implied by 
the Starobinsky-Yokoyama formula~(\ref{relation mass vs numerical mass:2}).
At a second sight however~(\ref{physical mass around origin}) is not surprising at all;
it simply tells us that for small field amplitudes, 
\begin{equation} \rho_{\rm eff}
 \propto{\rm exp}\bigg(
                    -\frac{\phi_{\rm b}^2}{2\langle\phi^2\rangle_{\rm stoch}}
                  \bigg)
\;;\qquad (\phi_{\rm b}\ll \phi_0)
\,,
\nonumber
\end{equation}
which is precisely what one would expect from a stochastic theory
in its Gaussian limit! (Albeit the theory is far from being Gaussian, for 
weak fields it is approximately Gaussian.)

Let us now check whether our result~(\ref{mass at origin}) agrees
with the numerical findings in figure~\ref{fig:Veff-rhoeff}.
In the weak coupling limit when $\zeta\rightarrow\infty$, 
$M^2(0)\rightarrow 1$, while in the strong coupling limit 
($\zeta\rightarrow 0$), 
$M^2(0)\rightarrow [\Gamma(1/4)/\Gamma(3/4)]\sqrt{\zeta}$. 
For $\zeta = 0.1,1,10$, Eq.~(\ref{mass at origin}) 
predicts $\{0.72,1.20,1.03\}$, which agrees well
with the approximate curves on the left panel of figure~\ref{fig:Veff-rhoeff}.
These values are to be compared with the 
values implied by Eq.~(\ref{relation mass vs numerical mass:2}):
$M_{\rm SY}^2(0) \simeq \{1.3,6,77\}$ for $\zeta=\{0.1,1,10\}$. 
We have thus found that the Starobinsky-Yokoyama procedure typically predicts
a much larger mass term at the origin than what is implied by
the effective potential approach. 

 In order to get a better feeling for the coupling dependence of 
the physical mass~(\ref{physical mass around origin}), 
consider the week and strong coupling regimes of 
Eq.~(\ref{physical mass around origin}),
\begin{equation}
m^2(0)\stackrel{\lambda\rightarrow 0}{\longrightarrow} 
        \frac{\lambda H^4}{16\pi^2\mu^2}
\;;\qquad 
m^2(0)\stackrel{\lambda\rightarrow \infty}{\longrightarrow} 
        \frac{\Gamma(1/4)}{\Gamma(3/4)}\frac{\sqrt{\lambda} H^2}{8\pi}
          \simeq 0.1177 \sqrt{\lambda} H^2
\,.
\label{physical masses: weak and strong limit}
\end{equation}
Thus, the strong coupling behavior is qualitatively the same as 
that of Starobinsky-Yokoyama~\cite{Starobinsky:1994bd}
 $m^2(0)\sim \sqrt{\lambda} H^2$ (albeit the dimensionless prefactors do not agree),
but in the weak coupling regime the results are qualitatively different. 
The latter result in Eq.~(\ref{physical masses: weak and strong limit})
should not surprise us: in the strong coupling regime 
the curvature around the origin ($-\mu^2$) is negligible, and one recovers 
the expected result $m^2(0)\sim\sqrt{\lambda}H^2$ (which is 
the loop counting parameter of the resummed perturbation theory).
On the other hand, the weak coupling result 
in~(\ref{physical masses: weak and strong limit}) is quite surprising,
in that it tells us that one almost recovers the na\"\i ve dependence on the 
coupling constant, $m^2\propto \lambda H^2$ (the additional factor 
$H^2/\mu^2$ is not easily explicable). What is interesting about that 
result is that, in the limit when $\lambda\rightarrow 0$
and/or $\mu\rightarrow \infty$, small amplitude 
fluctuations at the timelike asymptotic infinity of de Sitter behave as 
massless field fluctuations, with the caveat that  
the amplitude of these fluctuations must 
satisfy $|\phi_{\rm b}|< \phi_0 = \mu\sqrt{6/\lambda}$,
which is the scale at which the $\phi_{\rm b}^4$ term in the effective 
potential kicks in.

 Next we consider the large field behavior of the effective 
potential~(\ref{Veff:O(1)}). In the Appendix 
we show that the asymptotic effective potential
is of the form,
\begin{eqnarray}
 \frac{8\pi^2}{3H^4}V_{\rm eff}(\phi_{\rm b})
    &=& \zeta\bigg[\Big(\frac{\phi_{\rm b}}{\phi_0}\Big)^4
        -2\Big(\frac{\phi_{\rm b}}{\phi_0}\Big)^2\bigg]
   + \frac{8\pi^2}{3H^4}\Delta V_{\rm eff}(\phi_{\rm b})
\nonumber\\
\frac{8\pi^2}{3H^4}\Delta V_{\rm eff}(\phi_{\rm b}) &\approx& 
  \frac{\zeta}2 +\frac12\ln\bigg(\frac{3\pi\zeta\phi_{\rm b}^2}{2\phi_0^2}\bigg) 
   +\ln\bigg(I_{\frac14}\Big(\frac{\zeta}{2}\Big)
        + I_{-\frac14}\Big(\frac{\zeta}{2}\Big)\bigg) 
    \;,\qquad (\phi_{\rm b}\gg \phi_0)
\,,
\label{Veff:asymptotic}
\end{eqnarray}
or equivalently,
\begin{eqnarray}
V_{\rm eff}(\phi_{\rm b})
    &=& -\frac{\mu^2}{2}\phi_{\rm b}^2+\frac{\lambda}{4!}\phi_{\rm b}^4
   + \Delta V_{\rm eff}(\phi_{\rm b})
\nonumber\\
\Delta V_{\rm eff}(\phi_{\rm b}) &\approx& \frac{3\mu^4}{4\lambda}
 + \frac{3H^4}{16\pi^2}\ln\bigg(\frac{\phi_{\rm b}^2}{H^2}\bigg) 
 +\frac{3H^4}{8\pi^2}
    \ln\bigg[\frac{\pi^{3/2}\mu}{H}
             \bigg(I_{\frac14}\Big(\frac{\zeta}{2}\Big)
            +I_{-\frac14}\Big(\frac{\zeta}{2}\Big)
       \bigg)\bigg] 
    \;,\qquad (\phi_{\rm b}\gg \phi_0)
\,,
\label{Veff:asymptotic:2}
\end{eqnarray}
Firstly, the potential~(\ref{Veff:asymptotic}--\ref{Veff:asymptotic:2})
fits excellently the numerical $V_{\rm eff}$ for large $\phi_{\rm b}$, 
as can be seen in the left panel of figure~\ref{fig:Veff-rhoeff}
(dotted curves). Secondly, the leading term ($\zeta(\phi_{\rm b}/\phi_0)^4$)
and the subleading term ($-2\zeta(\phi_{\rm b}/\phi_0)^2$)
are the same as in the tree level potential~(\ref{rho V1}).
However, there is also the contribution
$\Delta V_{\rm eff}$ that originates from integrating out fluctuations on all 
super-Hubble scales,
and it consists of a logarithmic contribution 
($(1/2)\ln[(\phi_{\rm b}/\phi_0)^2]$) and a field independent contribution. 
These contributions are felt by large field excitations as a result of
integrating out fluctuations on all scales, 
and they introduce an average upward shift in the effective potential
that grows weakly (logarithmically) with the field amplitude.
This means that large amplitude, deep infrared fields in de Sitter 
are not distributed according to the Coleman-Weinberg effective 
potential~\cite{Coleman:1973jx}.
A comparison with the Starobinsky-Yokoyama
effective potential~(\ref{effective potential: Starobinsky-Yokoyama})
reveals that, in the large amplitude limit,
it recovers the leading (quartic) contribution in the field correctly, but it
fails at lower orders (quadratic, {\it etc}), and hence it does not represent
a very good fit, as can be seen on the left panel in
figure~\ref{fig:Veff-rhoeff} (for $\zeta = 10$ long dashed green).
In order to get a better feeling for $\Delta V_{\rm eff}$ in~(\ref{Veff:asymptotic:2})
notice that in the weak and strong coupling case it reduces to, 
\begin{equation}
\Delta V_{\rm eff}(\phi_{\rm b})
   \;\;\stackrel{\lambda\rightarrow 0}{\longrightarrow} \;\; 
\frac{3\mu^4}{2\lambda}
  +\frac{3H^4}{16\pi^2}\ln\left(\frac{\lambda\phi_{\rm b}^2}{\mu^2}\right)
\;;\qquad 
\Delta V_{\rm eff}(\phi_{\rm b})
   \;\;\stackrel{\lambda\rightarrow \infty}{\longrightarrow} \;\;
\frac{3H^4}{16\pi^2}
   \ln\left(\frac{\pi^2\sqrt{\lambda}\phi_{\rm b}^2}{\Gamma^2(3/4)H^2}\right)
\,.
\label{V0 expansions}
\end{equation}
This means that the contribution from integrated fluctuations grows 
as $\lambda$ decreases. In fact there is an upper bound for $\Delta V_{\rm eff}$
which comes from the natural assumption, $\mu,\phi_0  \lesssim m_{\rm p}$, 
which then implies that  
$\Delta V_{\rm eff} \lesssim m_{\rm p}^4/4$, which is of the order of
the Planck energy density. In the strong coupling regime however, 
$\Delta V_{\rm eff}$ remains limited to $\Delta V_{\rm eff}\sim H^4$.
 
  Next, we consider the backreaction on the geometry, for which we need 
the energy stored in the fluctuations.
According to the Starobinsky-Yokoyama prescription, 
the energy density in stored in quantum fluctuations is 
\begin{equation}
  \langle V\rangle_{\rm stoch} =  -\frac{\mu^2}{2}\langle \phi^2\rangle_{\rm stoch}
                + \frac{\lambda}{4!}\langle\phi^4\rangle_{\rm stoch}
\,,
\label{rho:SY}
\end{equation}
where $\langle \phi^2\rangle_{\rm stoch}$ and $\langle \phi^4\rangle_{\rm stoch}$
are given in Eqs.~(\ref{expec value O(1)}) and~(\ref{phi2n:stoch}).
For $\zeta = \{0.1,1,10\}$  Eq.~(\ref{rho:SY}) gives
$[8\pi^2/(3H^4)]\langle V\rangle_{\rm stoch} = \{0.11,-0.58,-9.48\}$. 
On the other hand, we can numerically evaluate $\langle V_{\rm eff}\rangle$ 
and we get for $\zeta = \{0.1,1,10\}$,
$[8\pi^2/(3H^4)]\langle V_{\rm eff}\rangle_{\rm stoch} 
  = \{0.39,0.34,0.23\}$, respectively. 
We thus see that the na\"\i ve contribution from quantum stochastic fluctuations
as given by the Starobinsky-Yokoyama procedure starts as a positive constant 
($=1/4$) when $\zeta\rightarrow 0$, but then it becomes negative as $\zeta$ increases,
approaching $[8\pi^2/(3H^4)]\langle V\rangle_{\rm stoch}\simeq -2\zeta$ 
when $\zeta\rightarrow \infty$.
This is the energy density perceived by the fields that vary over the Hubble scale,
but cannot be used for backreaction on the background geometry, for which we need
the energy density perceived by very deep infrared modes,
which is just $\langle V_{\rm eff}\rangle_{\rm stoch}$.
The contribution $[8\pi^2/(3H^4)]\langle V_{\rm eff}\rangle_{\rm stoch}$
is (a) positive for all values of the coupling constants $\zeta$ and 
(b) it decreases as $\zeta$ increases.
This is so because, when $\lambda\rightarrow 0$ and close to the origin
where $\rho$ is significantly different from zero, $V_{\rm eff}\simeq 0$ and it 
is very flat, and where
$V_{\rm eff}$  is large ($\phi_{\rm b}\gg \phi_0$),
$\rho$ shoots rapidly to zero. 
The energy $\langle V_{\rm eff}\rangle_{\rm stoch}$ is to be added as 
a {\it positive backreaction} $(8\pi G_N)\langle V_{\rm eff}\rangle_{\rm stoch}$ 
to the background cosmological constant $\Lambda_0$, 
effectively increasing the rate of Universe's expansion.
Since $[8\pi^2/(3H^4)]\langle V_{\rm eff}\rangle_{\rm stoch}$ is not greater than 
unity, the contribution to the cosmological constant is of the order 
$\Delta\Lambda = (8\pi G_N)\langle V_{\rm eff}\rangle_{\rm stoch}
       = (8\pi G_N){\cal O}(H^4)$,
which implies $\Delta\Lambda/\Lambda_0 = {\cal O}(H^2/m_{\rm p}^2)$
and hence it is typically small.

 Next, we shall consider the entropy stored in fluctuations at a point $\vec x$
at asymptotic timelike future infinity $i^+$.
Since stochastic formalism does not contain all of the information necessary to 
reconstruct the von Neumann 
entropy of the state~\cite{Weenink:2011dd,Koksma:2010dt,Koksma:2010zi,Prokopec:2006fc},
we resort to the less fundamental, but simpler, concept of
the Shannon (or Gibbs) entropy~\cite{Weenink:2011dd}, which is defined as 
$S_{\rm Shannon}=-\langle\ln(\rho\mu_0)\rangle_{\rm stoch}$, which when 
adapted to our problem yields, 
\begin{equation}
 S_{\rm Shannon, SY}=\frac{8\pi^2}{3H^4}\langle V\rangle_{\rm stoch} 
          + \ln\Big(\frac{Z_1}{\mu_0}\Big) 
\,,
\label{Shannon entropy}
\end{equation}
which in the strong and weak coupling limit gives
(for simplicity we set the scale $\mu_0=\phi_0$),
\begin{equation}
S_{\rm Shannon, SY}\stackrel{\zeta\rightarrow 0}{\longrightarrow}
       \ln\Big(\frac{\zeta}{4}\Big) +\frac14 
   + \ln\left(\frac{\pi}{\sqrt{2}\Gamma(3/4)}\right)
\;;\qquad  
S_{\rm Shannon, SY}\stackrel{\zeta\rightarrow \infty}{\longrightarrow} 
   -\frac12\ln\Big(\frac{\zeta}{\pi}\Big)
\,.
\nonumber
\end{equation}
The fact that in the weakly coupled regime $S_{\rm Shannon, SY}<0$ 
 should not be of a concern, since $S_{\rm Shannon, SY}$ in~(\ref{Shannon entropy})
is defined up to a constant determined by a mass scale $\mu_0$, and $\mu_0$
can be chosen such to keep $S_{\rm Shannon, SY}>0$.
In particular, for $\zeta=\{0.1,1,10\}$ one gets,  
$S_{\rm Shannon, SY}=\{1.52,1.09,-0.033\}$.
This is to be compared with the Shannon entropy implied by the effective theory, 
$S_{\rm Shannon, eff}=-\langle \ln(\rho_{\rm eff}\mu_0)\rangle_{\rm stoch}$,
which yields $S_{\rm Shannon, eff}=\{1.375,0.983,0.812\}$ 
for $\zeta=\{0.1,1,10\}$ and $\mu_0=\phi_0$, respectively.
Of course, $S_{\rm Shannon, eff}>0$ for arbitrary coupling constant provided 
$\mu_0$ is chosen large enough ($\mu_0=\phi_0$ suffices). This is so simply because 
$V_{\rm eff}>0$ ($\forall \phi_{\rm b},\zeta>0$). Again, we have seen that 
the Shannon entropy of the effective field theory gives more reasonable results 
than the corresponding quantity calculated by the Starobinsky-Yokoyama procedure.

 Finally, we give one last cursory look at the $V_{\rm eff}$ shown in 
figure~\ref{fig:Veff-rhoeff}, and observe that in the weak coupling limit 
(see the $\zeta = 10$ curves on the left panel of figure~\ref{fig:Veff-rhoeff})
the effective potential $V_{\rm eff}$ become flatter around the origin 
($\phi_{\rm b}<\phi_0$) as $\zeta$ becomes larger, and it becomes steeper
for large values of the field ($\phi_{\rm b}>\phi_0$), such that there is a 
sudden change in the effective potential curvature at $\phi_{\rm b}\sim \phi_0$. These 
features are reminiscent of a Maxwell construction
(see {\it e.g.} Ref.~\cite{Peskin:1995ev}),
which is an approximate procedure for constructing the free energy/effective potential
close to a critical point (where fluctuations become massless and
long range correlations develop). Indeed, in the weak coupling regime 
one can approximate the effective potential by the following
Maxwell-like construction: use Eq.~(\ref{Veff:origin}) for $\phi_{\rm b}<\phi_0$
and Eqs.~(\ref{Veff:asymptotic}--\ref{Veff:asymptotic:2}) 
for $\phi_{\rm b}>\phi_0$, and this will give a reasonable approximation
to the true effective potential, that exhibits a discontinuity in curvature at 
$\phi_{\rm b}\simeq\phi_0$.

\subsection{The $O(N)$ model}
\label{The O(N) model:2}

In analogy to the convexity condition in the $O(1)$ model~(\ref{derivatives1}),
in this more general case the effective potential $V_{\rm eff}(\vec\phi^{\,\rm b})$ 
($\vec\phi^{\,\rm b}=(\phi^{\rm b}_i, i=1,..,N)$) 
will be convex provided the corresponding Hessian matrix
\begin{eqnarray}
\dfrac{\partial^2 V_{\rm eff}(\vec{\phi}^{\,\rm b})}
  {\partial\phi^{\rm b}_i\partial\phi^{\rm b}_j}
   = \dfrac{\partial J_{i}(\vec \phi^{\,\rm b})}{\partial \phi^{\rm b}_j}
\label{Hessian matrix}
\end{eqnarray}	
is strictly positive ({\it i.e.} all of its eigenvalues are strictly positive), 
where $\vec J=(J_i)$ is the source current 
($\rho_{\vec J}(\vec\phi\,)=\rho(\vec\phi\,){\rm e}^{\vec J\cdot \vec\phi}$).
Taking the inverse of the matrix~(\ref{Hessian matrix}) (which exists provided 
its determinant does not vanish) we get
({\it cf.} Eqs.~(\ref{convexity condition}--\ref{convexity condition:2})) 
\begin{eqnarray}
\left(\dfrac{ \partial^{2} V_{\rm eff}(\vec{\phi}^{\,\rm b}) }
                             {\partial \phi^{\rm b}_i\partial \phi^{\rm b}_j} \right)^{-1} 
   = \langle\phi_i\phi_j\rangle_{\vec J} 
    - \langle\phi_i\rangle_{\vec J}\,\langle\phi_j\rangle_{\vec J} 
\,.
\label{Hessian:O(N)}
\end{eqnarray}
The simplest way of finding the eigenvalues of the inverse Hessian
matrix~(\ref{Hessian:O(N)}) is to make 
use of the $O(N)$ symmetry, which implies that there exists an orthogonal transformation 
$R$ ($R\cdot R^T = I = R^T\cdot R$) 
which brings the source current $\vec J$ to the first component $\vec J^\prime = R\cdot \vec J$,
where $\vec J^\prime =(J,0,..,0)$ and $J=\|\vec J\,\|$. Now, making use of the invariance 
$\vec J^{\,T}\cdot \vec \phi = (\vec J^\prime)^{T}\cdot \vec \phi^\prime$, 
$\vec \phi^\prime =R\cdot \vec \phi$, and of the fact that  
the Jacobian of the transformation $\vec \phi\rightarrow \vec \phi^\prime$ equals unity,
we conclude that the partition function $Z(\vec J)=Z(\vec J^\prime)$
(this just means that $Z(\vec J\,)$ is a function of $\vec J^{\,2}$, which is invariant 
under orthogonal transformations). 
Consequently, we can act with $R^T$ from the left and with $R$ from the right 
on~(\ref{Hessian:O(N)}) to obtain a rotated Hessian matrix, 
$\langle \phi'_{i}\phi'_{j} \rangle_{\vec J^\prime}
  - \langle \phi'_{i} \rangle_{\vec J^\prime} \langle \phi'_{j}\rangle_{\vec J^\prime}$, 
which is in its diagonal form! 
In order to see that, we shall now write the positivity condition for
the nonvanishing elements of the rotated inverse Hessian matrix, 
\begin{equation}
\begin{cases}
  i=j=1:\langle (\phi_{1}')^{2} \rangle_{\vec J'}-  \langle \phi'_{1} \rangle_{\vec J'}^{2} > 0,
 \\
  i=j\neq 1 :\langle (\phi'_{i})^{2} \rangle_{\vec J'}>0 
\,,
\end{cases}
\end{equation}
where in the last inequality we used the fact that 
$\langle \phi'_{i}\rangle_{\vec J'}=0$ for $j=2,3,..,N$, which is a consequence of the fact 
that $J^\prime_i=0$ ($i=2,3,..,N$). Now the first inequality is proved identically 
as in the $O(1)$ case, while the second line inequality is trivially true.
This completes the proof that $V_{\rm eff}$ is convex, implying that 
deep infrared scalar fields ${\vec \phi}^{\,\rm b}$ on de Sitter
cannot undergo phase transitions.

 Needless to say, one can perform numerical and analytical analysis of the resulting 
$V_{\rm eff}(\vec \phi^{\,\rm b}\,)$ by using analogous methods as in 
sections~\ref{The O(1) model} 
and~\ref{Analytical approximations for the effective potential}.


 \section{Discussion}
 \label{Discussion}

We use the formalism of Starobinsky's stochastic inflation to show in 
section~\ref{Symmetry restoration} that there can be no scalar field condensate formation in an
$O(N)$ symmetric scalar field theory. For simplicity we have separately considered
the $O(1)$ model (a real scalar field) and the $O(2)$ model (a complex scalar field). 
Nevertheless, as figure~\ref{fig:initial stochastic PDF} clearly indicates,
the probability distribution function for stochastic field
that varies over the Hubble scale shows a double-peak structure,
implying that first order transitions can exist 
during inflation, such that one can foresee formation of topological defects on 
sub-Hubble scales that persist on super-Hubble scales, at least for a while.

 In section~\ref{Effective potential} we borrow the concept of effective field theories 
to study the probability distribution of the deep infrared fields in de Sitter space,
that `live' on the asymptotic future timelike infinity $i^+$ of de Sitter.
We prove that the effective potential that governs the probability distribution
of deep infrared fields of an $O(N)$ scalar field theory must be {\it strictly convex}. 
Our proof of convexity holds for 
arbitrary integer $N$ and for arbitrary values of the background field. 
Our proof applies for an arbitrary quartic (Hubble scale renormalized) potential, 
and we strongly suspect that, by making use of analogous methods, one can construct
a proof of convexity for an arbitrary $O(N)$ symmetric local scalar field theory. 
An important consequence of this result is the impossibility 
of field condensation at the asymptotic future timelike infinity $i^+$ of de Sitter space
illustrated in figure~\ref{fig:CarterPenrose-dS}. 
The physical reason for this symmetry restoration
is the strong infrared vacuum fluctuations on de Sitter space. 

 It is worth mentioning that the effective potential
can be fully characterized in terms of just one physical dimensionless parameter
$\zeta$ defined in~(\ref{rho VN}), which characterizes the inverse coupling strength.
Furthermore, while for small background field values the effective potential is Gaussian 
(corresponding to a positive mass parameter~\cite{Starobinsky:1994bd,Beneke:2012kn})
for large field values it reduces, as expected, to the tree level quartic potential
plus a correction that logarithmically depends on the background field.
On the other hand, as our analysis in 
section~\ref{Analytical approximations for the effective potential} shows, 
the curvature of $V_{\rm eff}$ is much softer than the one implied by 
the Starobinsky-Yokoyama approach.

 In section~\ref{Analytical approximations for the effective potential} we also discuss
the backreaction on the background geometry, and show that
 it is always positive and suppressed independently on the choice of parameters.
The backreaction is typically of the order
$\Delta\Lambda/\Lambda = {\cal O}(\Lambda/m_{\rm p}^2)$,
which is to be contrasted with the result obtained by
the Starobinsky-Yokoyama procedure, which can be either positive or negative (in the limit
when $\zeta\rightarrow\infty$, $\langle V\rangle_{\rm stoch}\rightarrow -2\zeta$).

 Since astronomical observations of cosmic microwave background anisotropies
and Universe's large scale structure measure some fixed finite physical
scales, it would be of interest to extend the analysis in
this paper to study a real scalar field probability distribution, and 
the corresponding field correlators,
on some finite physical scale $\mu\ll H$. In order to properly 
study these, one would have to develop a full effective theory 
for the probability distribution $\rho_{\rm eff}(\phi,\mu,t)$ on some finite scale $\mu$, 
for which a first -- but possibly na\"\i ve~\cite{Janssen:2009pb,Bilandzic:2007nb}
-- guess is given by Eq.~(\ref{Fokker-Planck:RG}).  
Since $\rho_{\rm eff}(\phi,\mu,t)$ is the classical equivalent of the density operator of 
quantum field theory,  $\rho_{\rm eff}(\phi,\mu,t)$ must contain complete information 
about the theory, that includes (equal time) field correlators, 
which are of relevance for cosmology.

 Finally, it would be of interest to extend our analysis to other stochastic theories whose 
evolution is of the Langevin type. Interesting examples include the infrared dynamics 
of thermal field theories such as 
non-Abelian gauge theories~\cite{Bodeker:1998hm,Zahlten:2008im}.

\section*{Appendix}

\subsection*{O(N) case}

 Here we show that the function $F(\zeta,N)$ in
Eq.~(\ref{O(N):symmetry breaking:condensate:4})
is strictly negative for any positive integer $N$ and real $\zeta>0$. 
Let us begin by recalling the series around $z=0$
of the confluent hypergeometric function 
(which has infinite radius of convergence):
\begin{equation}
{}_1F_1(a;b;z)=\sum _{n=0}^\infty\frac{\Gamma(a+n)\Gamma(b)}{\Gamma(a)\Gamma(b+n)}
               \frac{z^n}{n!}
\,,
\label{Confluent hypergeometric function}
\end{equation}
where $\Gamma(z)$ denotes the gamma function, $\Gamma(z+1)=z\Gamma(z)$.
It is now convenient to rewrite  $NF(\zeta,N)=A(\zeta,N)/B(\zeta,N)$, where:
\begin{equation}
 B=\Gamma\left(\frac{N}{4}\right)\times {}_1F_1\left(\frac{N}{4};\frac{1}{2};\zeta\right)
   +2 \sqrt{\zeta}\;\Gamma\left(\frac{N\!+\!2}{4}\right)
       \times{}_1F_1\left(\frac{N\!+\!2}{4};\frac{3}{2};\zeta\right)
\label{App:B}
\end{equation}
and 
\begin{eqnarray}
A &=& N\Gamma\left(\frac{N}{4}\right)\times{}_1F_1\left(\frac{N}{4};\frac{1}{2};\zeta\right)
+ 2N\sqrt{\zeta}\Gamma\left(\frac{N\!+\!2}{4}\right)
        \times{}_1F_1\left(\frac{N\!+\!2}{4};\frac{3}{2};\zeta\right)
\nonumber\\
&-& 2(N\!+\!2)\Gamma\left(\frac{N}{4}\!+\!1\right)
      \times {}_1F_1\left(\frac{N}{4}\!+\!1;\frac{3}{2};\zeta\right)
  -\frac{N\!+\!2}{\sqrt{\zeta}}\Gamma\left(\frac{N\!+\!2}{4}\right)
   \times {}_1F_1\left(\frac{N\!+\!2}{4};\frac{1}{2};\zeta\right)
\,.\quad
\label{App:A}
\end{eqnarray}
Let us first consider $B(\zeta,N)$ in Eq.~(\ref{App:B}).
In order to show that $B>0$ for any $\zeta>0$ and for any positive integer $N$,
observe first that the confluent hypergeometric function ${}_1F_1(a;b;z)>0$
if $a>0$, $b>0$ and $z>0$. This is so simply because 
all of the coefficients in the series~(\ref{Confluent hypergeometric function}) 
contain gamma functions of positive arguments, which are strictly positive.
Now, because Eq.~(\ref{App:B}) contains confluent hypergeometric functions
with positive indices $a,b$ and of a positive argument $\zeta>0$, it follows that 
$B(\zeta,N)>0$ for any $\zeta>0$ and any positive integer $N$.

 To complete the proof we need to show that the function $A$ in~(\ref{App:A})
is strictly negative. In order to show that, observe that 
one can combine the first with the third and the second with the fourth term, to obtain,
\begin{eqnarray}
A & =& -\frac{4}{\sqrt{\zeta}}\Gamma\Big(\frac{N}{4}+\frac32\Big)
  -\sum_{n=0}^\infty \frac{\Gamma\Big(\frac{N}{4}+n\Big)\Gamma\Big(\frac32\Big)}
                          {\Gamma\Big(\frac{3}{2}+n\Big)}
    \frac{\zeta^n}{n!}\bigg[\frac{N^2}{2}+4n\bigg]
\nonumber\\
&& -\, \sqrt{\zeta}\sum_{n=0}^\infty 
      \frac{\Gamma\Big(\frac{N}{4}+\frac12+n\Big)\Gamma\Big(\frac32\Big)}
                          {\Gamma\Big(\frac{3}{2}+n\Big)}
    \frac{\zeta^n}{(n+1)!}\bigg[\frac{N^2}{2}+4n+2\bigg] <0
\,,
\label{App:A2}
\end{eqnarray}
which is strictly negative since all three terms are negative for arbitrary $\zeta>0$ and 
positive integer $N$, completing the proof.
Notice that the first term in~(\ref{App:A2}) originates from the first term in the series 
of the last hypergeometric function in~(\ref{App:A}),
and that all gamma functions appearing in~(\ref{App:A2}) are strictly positive.

\subsection*{Asymptotic effective potential}
\label{Asymptotic effective potential}

 In this appendix we present a derivation of
the effective potential~(\ref{Veff:asymptotic})
for large field values $\phi_{\rm b}\gg \phi_0$. 
In doing so we shall make use of 
the formulae~(\ref{Z(J)}--\ref{rho_eff}) 
and~(\ref{Z(J):2}--\ref{phi2n:stoch}) 
in the main text.
Since the mapping $J\mapsto \phi_{\rm b}$ is a monotonically 
increasing function, when $\phi_{\rm b}$ is large so is $J$.
When summing the series~(\ref{Z(J):2}), observe that in the limit when 
$J$ is large, terms with a large $n$ dominate the sum, and hence we need 
a large index expansion of the confluent hypergeometric function,
which can be, for example, found in Eqs.~(9.228) and~(9.220.3) 
of Ref.~\cite{Gradshteyn:1965},
\begin{equation}
 {}_1F_1(\alpha;\beta;z)\;\;\stackrel{|\alpha|\rightarrow\infty}{\sim}\;\;
  \frac{\Gamma(\beta){\rm e}^{z/2}z^{\frac{1-2\beta}{4}}}{\sqrt{\pi}}
      \times\cos\bigg(2\sqrt{\Big(\frac{\beta}{2}-\alpha\Big)z}
               +\Big(\frac14-\frac{\beta}{2}\Big)\pi\bigg)
\,,
\label{1F1:asymptotic}
\end{equation}
from which we infer,
\begin{equation}
  {}_1F_1\Big(\frac{n}{2}+\frac14;\frac12;\zeta\Big)
    \;\;\stackrel{n\rightarrow\infty}{\sim}\;\;
      {\rm e}^{\zeta/2}\cosh\big(\sqrt{2n\zeta}\,\big)
\;;\qquad 
  {}_1F_1\Big(\frac{n}{2}+\frac34;\frac32;\zeta\Big)
    \;\;\stackrel{n\rightarrow\infty}{\sim}\;\;
      {\rm e}^{\zeta/2}\frac{\sinh\big(\sqrt{2n\zeta}\,\big)}{\sqrt{2n\zeta}}
\,.
\label{1F1:asymptotic:2}
\end{equation}
Inserting this into~(\ref{Z(J):2}--\ref{phi2n:stoch}) yields
\begin{eqnarray}
Z(j) &\equiv& {\rm e}^{-W(j)} = \sum_{n=0}^\infty {\rm e}^{-\omega(n,\zeta,j)}
\label{1F1:asymptotic:3}
\\
\omega(n,\zeta,j) &=& -\sqrt{2n\zeta}
  -\frac{n}2\ln\bigg(\frac{j^4}{\zeta}\bigg)
  -\ln\Bigg[\frac{\Gamma\big(\frac{n}2\!+\!\frac14\big)
          \!\!+\sqrt{\frac2n}\Gamma\big(\frac{n}2\!+\!\frac34\big)}
                {\Gamma(2n\!+\!1)}\Bigg]
  +\ln\bigg[2\pi\zeta^{1/4}\bigg(I_{\frac14}\Big(\frac{\zeta}{2}\Big)
          +I_{-\frac14}\Big(\frac{\zeta}{2}\Big)\bigg)\bigg]
\,,
\nonumber
\end{eqnarray}
where we have used a rescaled (dimensionless) current $j=\phi_0 J$
(similarly, we shall be using a dimensionless background field 
$\varphi_{\rm b} = \phi_{\rm b}/\phi_0$).
The sum~(\ref{1F1:asymptotic:3}) is hard to evaluate. However, by noticing 
that in the limit of large $j$, the sum is dominated 
by large $n$'s, we can replace the sum by an integral, 
$\sum_{n=0}^\infty\rightarrow \int_0^\infty dn$,
and make a saddle point approximation to the integral.
This means that we can use a large $n$ expansion 
of $\omega(n,\zeta,j)$, and approximate it by its 
expansion around the stationary point $n_0$, where 
$\partial \omega/\partial n = 0$,
\begin{equation}
 \omega(n,\zeta,j)\approx \omega_0(\zeta,j) 
+ \frac{1}{2}\omega^{\prime\prime}_0(\zeta,j)(n-n_0)^2
\,,
\label{saddle point approximation}
\end{equation}
where (in the limit when $n\rightarrow \infty$),
\begin{equation}
 n_0 \approx \frac12\left(\frac{j^4}{4\zeta}\right)^{1/3}
        \left[1 +\frac13\left(\frac{4\zeta}{j}\right)^{2/3}
                -\left(\frac{4\zeta}{j^4}\right)^{1/3}\right]
\label{n0}
\end{equation}
The resulting Gaussian integral is convergent if
$\omega^{\prime\prime}_0(\zeta,J)>0$,
which is indeed the case, and it evaluates to,
\begin{equation}
 Z(j)={\rm e}^{-W(j)}\approx {\rm e}^{-\omega_0}
   \sqrt{\frac{2\pi}{\omega_0^{\prime\prime}}}
   \left[1-\frac12{\rm erfc}\Big(\sqrt{\omega_0^{\prime\prime}/2}\,\times n_0\Big)\right]
\label{saddle point integral}
\end{equation}
where the complement error function ${\rm erfc}(x)=1-{\rm erf}(x)$ 
can be neglected when the argument is large because, in the large 
argument limit, ${\rm erfc}(x)\sim {\rm e}^{-x^2}\!/[\sqrt{\pi}x]$.
The standard expression now yields the following approximate expression
for $\varphi_{\rm b}$,
\begin{equation}
 \varphi_{\rm b} = \frac{\partial\ln[Z(j)]}{\partial j}
       \approx \left(\frac{j}{4\zeta}\right)^{1/3}
\left[1+\frac{1}{3}\left(\frac{4\zeta}{j}\right)^{2/3}
          -\frac13\left(\frac{4\zeta}{j^4}\right)^{1/3}\right]
\,,
\label{phib vs J}
\end{equation}
which when inverted gives,
\begin{equation}
j \approx 4\zeta \varphi_{\rm b}^3  
  \left[1 -\frac{1}{\varphi_{\rm b}^2} + \frac{1}{4\zeta\varphi_{\rm b}^4}
  \right]
\,.
\label{J vs phib}
\end{equation}
Inserting this into Eq.~(\ref{Veff:O(1)}) and making 
use of~(\ref{saddle point integral}) yields
the asymptotic effective potential
\begin{equation}
 \frac{8\pi^2}{3H^4}V_{\rm eff}(\phi_{\rm b})
    \approx \zeta\Big(\varphi_{\rm b}^4-2\varphi_{\rm b}^2+\frac12\Big)
    +\frac12\ln\bigg(\frac{3\pi\zeta\varphi_{\rm b}^2}{2}\bigg) 
   +\ln\bigg(I_{\frac14}\Big(\frac{\zeta}{2}\Big)
        + I_{-\frac14}\Big(\frac{\zeta}{2}\Big)\bigg) 
    \;,\qquad (\varphi_{\rm b}\gg 1)
\,,
\label{Veff:asymptotic:appendix}
\end{equation}
which is used in section~\ref{Analytical approximations for the effective potential},
see Eq.~(\ref{Veff:asymptotic}).

\section*{ACKNOWLEDGMENTS}

We would like to thank Dra\v zen Glavan and Frits Beukers from Utrecht university
for their important hints.

\end{document}